# Made to measure: an introduction to quantification in microscopy data

Siân Culley[1], Alicia Cuber Caballero[1], Jemima J Burden[2], Virginie Uhlmann[3]

[1]Randall Centre for Cell and Molecular Biophysics, King's College London, London, UK
[2]Laboratory for Molecular Cell Biology, UCL, London, UK
[3]European Bioinformatics Institute (EMBL-EBI), EMBL, Cambridge, UK

## Abstract

Images are at the core of most modern biological experiments and are used as a major source of quantitative information. Numerous algorithms are available to process images and make them more amenable to be measured. Yet the nature of the quantitative output that is useful for a given biological experiment is uniquely dependent upon the question being investigated. Here, we discuss the 3 main types of visual information that can be extracted from microscopy data: intensity, morphology, and object counts or categorical labels. For each, we describe where they come from, how they can be measured, and what may affect the relevance of these measurements in downstream data analysis. Acknowledging that what makes a measurement "good" is ultimately down to the biological question being investigated, this review aims at providing readers with a toolkit to challenge how they quantify their own data and be critical of conclusions drawn from quantitative bioimage analysis experiments.

## Introduction

Quantification is at the heart of modern biological research, and image data is one of the major sources of quantitative information[1–3]. The past decade has seen an explosion of software tools dedicated to bioimage analysis[4]. These methods most often adapt computer vision algorithms to the specificities of biological image data, such as poor contrast and the presence of complex objects, with the aim to speed up and streamline the quantification process[5]. Nowadays, visual information held in microscopy images can be processed and quantified by computer programs in a largely automated manner, opening up the possibility to analyse data at scale, in a reproducible and objective manner[6].

While greatly enabling biological research, the democratisation of quantitative image analysis tools also poses some challenges. One of the biggest challenges is identifying which of the dozens of quantifications that can be generated by these tools are most informative and relevant to the biological question being studied. To navigate this, a helpful strategy is to scrutinise all steps from experimental design to image acquisition and ultimately data processing and analysis. This process allows each experimental stage to be tailored to best inform the hypothesis being investigated, but can also highlight aspects that may adversely influence the quantitative output. The first step consists of defining the goal of the imaging experiment (what quantitative property is at the centre of the question being investigated) prior to acquiring data, and using this information to guide experimental design[7]. Once data starts being generated, the focus shifts to the many experimental factors and acquisition parameters affecting the quality of imaging. These can include sample and labelling properties, such as photobleaching and cross-talk, as well as hardware specifics[8,9]. Whenever appropriate imaging parameters are identified and used throughout an experiment, ensuring that they are appropriately recorded as metadata is crucial to ensure that downstream analysis is carried out in a meaningful way[10]. The same considerations apply to any image processing applied post-acquisition, such as methods to remove noise or improve contrast/resolution. After optimising experimental design, acquisition parameters, post-processing, and recording metadata, one is left with analysing the acquired image data. At this point, the last remaining step is to outline the best way to pull out the desired measurements. This raises questions about which aspects of visual data can be measured, which methods are available to do so, and what are their limitations.

Several excellent review papers assess the performance of image analysis methods in a benchmarking setting, meaning that methods are evaluated on the one general task they were designed to solve. General tasks, such as segmentation[11,12] and denoising[13], are however generally not the end goal of an experiment. The quantitative output needed to explore a biological question is indeed rarely to improve the quality of an image or partition it. Instead, segmentation and denoising are examples of operations that enable the final quantification task, which is usually to measure a specific phenotype. To complement the existing literature on image analysis algorithms, this review focuses on the quantification problem. We review common categories of quantitative readouts that can be extracted from visual data and the appropriate metrics to do so. We identify three main categories of quantitative information: image intensity, morphology, and counts or labels. For each of these categories, we describe the process that leads to this information being available in image data. We then review how that type of quantitative information can be extracted and interpreted, and discuss aspects impacting its quantification. We close the paper with a discussion on quality control and confidence. Considering the breadth of the topic, we choose to limit the scope of this paper to individual 2D images and 3D volumes, acquired with either fluorescence microscopy or electron microscopy (EM). We therefore do not cover measurements that are specific to time-series, multi-channel imaging, or to data obtained with specialised imaging modalities such as super-resolution microscopy or single particle electron microscopy.

In order to be able to formulate what one *wants* to measure, one must first understand what *can* be measured. By providing an overview of the big categories of quantitative measurements in image analysis, the goal of this review is to provide life science researchers with a framework to appreciate and scrutinise their own image data. We also aim to give insights on aspects of an experiment that impact the relevance of measurements in downstream analysis, and therefore enable readers to be critical about whether the conclusions of studies involving image quantification are meaningful or not.

## Image Intensity

When acquiring microscopy data one of the first things a researcher checks, either via visual assessment of the image or digital inspection of pixel values, is the intensity. But what do the pixel intensity values in microscopy images actually mean, and where do they come from?

### Generating contrast and capturing intensity

The image generation process fundamentally differs across imaging modalities. Contrast is most often induced by biochemical labels that are artificially introduced in samples specifically for imaging. Contrast can also be obtained in a label-free manner with dedicated optical components, as is the case for phase contrast and Differential Interference Contrast (DIC) imaging[14].

In fluorescence microscopy, molecules of interest are labelled with fluorescent species such as organic dyes and fluorescent proteins. Regardless of the microscope used and downstream analysis performed, the manner in which labelling is performed must be taken into account to provide context to any quantification. For example, if non-endogenous fluorescent protein fusion constructs are being used, how closely do measured intensities reflect endogenous protein distributions[15]? If immunofluorescence techniques are being used, what factors are impacting the ratio between number of epitopes of interest and the number of fluorophores (e.g. primary and secondary antibody concentrations, antigen masking affinity, clonality of antibodies)[16]? Most fluorescence intensity quantifications made from images are of the fluorophores themselves, not directly of the biological molecules of interest. This should be taken into consideration when translating any results from fluorescence intensity quantification into biological conclusions. The microscope modality also affects how much out-of-focus fluorescence contaminates the in-focus fluorescence measurements (**Figure 1a-d**). With EM imaging, the mechanism of contrast generation varies significantly depending upon the type of experiment performed, as well as the type of microscope and detectors used (**Figure 1e-h**). For the majority of the different types of EM experiments, contrast is introduced into the sample as part of sample preparation. Sample contrast is enhanced by the introduction of heavy metals (eg. osmium, lead, uranium, gold, silver etc) which bind to lipids, proteins, carbohydrates etc via chemical reactions, whereby the method, sequential order of addition, temperature, time of incubation, etc can impact the process by which the contrast is incorporated into the sample[17].

The entire purpose of any microscope is to transmit the biological information contained within the sample to the detector. The pixel intensities in the acquired images will depend on a number of factors in the imaging process, and understanding these factors is important for contextualising quantitative intensity measurements and assessing their accuracy.

In fluorescence microscopy, the intensities in an image represent the number of photons emitted by excited fluorophores. The absolute number of photons emitted by fluorescent molecules in the sample is primarily determined by the intensity of the excitation illumination incident on the sample. Depending on the imaging modality used in fluorescence microscopy, intensity information may also arise from fluorescent sources in the sample other than the labelled structures in the focal plane. In techniques capable of optical sectioning, such as confocal microscopy (**Figure 1b**), two-photon microscopy, and TIRF (see Glossary in **Table 1**, **Figure 1d**), out-of-focus fluorescence does not reach the detector, whereas images acquired using widefield microscopy will contain out-of-focus intensities (**Figure 1a**). All fluorescence microscopy images may also contain intensity contributions from autofluorescence (endogenous fluorescent species present within the sample in the absence of intentional labelling). Confocal z-stacks are frequently projected into a single 2D image for visualisation and analysis (**Figure 1c**); a 'sum slices' or average intensity projection will retain intensity information, whereas a maximum intensity projection will produce sharper images but with intensities that do not correspond to the total amount of fluorescence below each pixel and thus shouldn't be used for intensity quantification.

Increasing illumination intensity, regardless of the source, usually results in an increase in the number of photons emitted by fluorescence. However, increasing excitation intensities can lead to non-linear saturation of emitted fluorescence (**Figure 2a, 2b**), and increase the rate at which permanent photobleaching of fluorophores occurs (**Figure 2c**). When fluorescence intensity measurements are important, inhomogeneity of the illumination across the field-of-view can also create unwanted variability. The 'flatness' of the illumination can be characterised, and corrected for further quantitative measurements; this can be done using a homogeneously fluorescent test sample[8,18] (**Figure 2d**), or via computational methods[19,20] without a test sample.

For EM, the sample preparation protocol, the type of experiment, microscope and detectors used all impact the information captured within the images obtained. In conventional transmission electron microscopy (TEM), images are collected of ultrathin sections (50-100nm thick) of a resin embedded, contrasted sample. The image is created by the detection of electrons that pass through the sample and reach the detection mechanism. Contrast is generated by localised heavy metals, introduced during the sample preparation steps, scattering electrons, preventing their transmission through the sample (**Figure 1e**). Scanning electron microscopes (SEM) are another common type of electron microscope used for biological investigations, where a focused beam of electrons is scanned across the sample and resulting secondary and/or backscattered electrons or x-rays are collected to generate an image. The most common approach is the collection of secondary electrons that have interacted with the surface of a sample that has been fixed, dehydrated, dried and the surface coated in a thin layer of metal. The resulting image is a view of the surface topology of the sample from a particular view point (unless rotational imaging and photometry is applied), with a large depth of field. Contrast is provided by the differential angles of the detector, source and how the electron beam interacts with the limiting shape of the sample (**Figure 1f**). Back-scattered electrons (BSE) can also be separately collected and mapped onto the sample and provide information about the sample's elemental composition (**Figure 1g**). BSE imaging has recently been exploited in the development of a collection of volume EM (vEM) techniques, where either arrayed sections or blocks of resin embedded, fixed, contrasted samples are automatically imaged (**Figure 1h**), generating a large 3D volume of ultrastructural data at nm resolution, across scales of 10's-100's of microns[21]. Regardless of the detection technique, it is important to be aware that working distance, magnification, accelerating voltage and probe size and current are just some of the parameters that impact the resulting images in terms of resolution, depth of field, focus and contrast.

The final acquired images are formed by binning (see Glossary in **Table 1**) the detected photons (fluorescence microscopy) or electrons (EM) into pixels. The pixel size in a microscopy image plays a critical role in determining what quantitative information can be retrieved. The physical distance that each pixel represents (the pixel size) is primarily determined by properties of the detection path (for cameras, the physical size of the pixels on the chip, and for point detectors, the scanning parameters), the magnification, and the numerical aperture of the microscope. As a result, the accuracy of both intensity and morphology measurements from the same biological structures varies depending on the magnification and numerical aperture (NA) of the microscope objective, as shown on **Figure 3**.

Per the Nyquist-Shannon sampling theorem, retaining the resolution of a continuous signal (i.e. the spatially varying distribution of photons/electrons incident on the detector) in discrete digital space (i.e. the pixels in the acquired image) requires sampling at at least double the frequency of the smallest resolvable feature[22]. When the Nyquist-Shannon theorem is applied to the two-dimensional nature of images, the theoretical pixel size for adequate sampling should in fact be ~2.8 times smaller than the resolving power of the microscope[23]. This sampling should be observed if very fine structures within the sample are to be measured and quantified, as larger pixel sizes will lead to a loss of information due to undersampling. In addition to spatially binning detected photons or electrons into pixels, detectors also convert the measured intensity into an integer number. This value depends on the intensity of emitted fluorescence or scattered electrons, as well as detector settings such as gains and offsets (see Glossary in **Table 1**). However, it also depends on the bit depth of the acquired images. Bit depth determines the range of values that can be digitally stored within a pixel; most microscopy data is acquired at 8-, 12-, or 16-bit depth. A pixel can only store a number within the range $0 \rightarrow (2^N-1)$, where N is the bit depth. The effect of bit depth on intensity information is somewhat analogous to the effect of pixel size on spatial information; higher bit depths provide higher 'sampling' of intensities, which can provide higher precision for quantitative measurements. Critically, measurements should not be made from any pixels having either the minimum (0) or maximum ($2^N-1$) value as this is likely to represent incomplete or 'clipped' information in the image; unless the image was acquired in a very low fluorescence microscopy regime, pixels of value 0 may in fact represent a range of different 'real' fluorescence intensities that are below the range of the detector settings, and pixels of value $2^N-1$ may represent a range of real fluorescence intensities that are saturating the detector.

Extracting and interpreting intensity measurements

When it comes to extracting an intensity-based measurement from a fluorescence microscopy image, be that from the raw data or after processing, it is important to remember that fluorescence intensity measurements are always comparative. Standalone measurements of pixel or object intensity in images are often meaningless; they must be reported in the context of some baseline condition such as the background intensity, or the intensity of a comparable object under a different biological condition. For such comparisons to be made accurately, it is critical that acquisition parameters such as illumination intensity, magnification, pixel dwell time (point detectors) or exposure time (cameras), and detector gains are recorded and ideally kept consistent between different images. Any image processing pipelines should be applied equivalently to each image, including the ones that do not look like they 'need' it.

For some biological measurements, it makes sense to work with the absolute fluorescence values in images, such as monitoring the expression of a GFP-tagged protein during successive cell divisions[24]. However, when aggregating results from different images, relative fluorescence intensities are often used so that results can be aggregated. On the whole, any absolute measurements of intensity from fluorescence microscopy modalities such as widefield and confocal microscopy are critically dependent on labelling, acquisition settings, and post-processing. If comparisons of fluorescence intensity are to be made between different images then these parameters should be as identical as possible in each case. In EM, contrast intensity in images is rarely absolutely quantified as routinely controlling the contrast incorporation process and calibrating the detection process is fraught with challenges.

Factors impacting intensity information

A vast range of image processing operations can be applied to raw images following acquisition. If quantitative intensity information is to be extracted following image processing, it is then important to understand how processing affects image intensity (**Figure 1c**, **Figure 4**). During image processing, there are occasions where the bit depth of an image is changed. This is typically when a mathematical operation is performed on the image that generates values that are beyond the range of the bit depth (for example, a negative number) or have a non-integer component. Intensity quantification is still valid when performed on images after increasing the bit depth, but no intensity quantifications should be made from images following conversion to a lower bit depth. This is because conversion to a lower bit depth requires a rescaling of pixel values so that they fit within the smaller range, which results in a loss of information from the image.

If intensity measurements are to be made following image processing, then it is important that the processed values are still linearly related to the number of fluorescent molecules present in a given region of the image (**Figure 4D**). Iterative deconvolution (see Glossary in **Table 1**) methods have been shown experimentally to be largely linear with respect to intensity, although this can be microscope-dependent[25] (**Figure 4, 'Deconvolution'**). An example of a non-linear image processing operation is the Super-Resolution Radial Fluctuations (SRRF) method[26] (**Figure 4, 'SRRF'**). This is an example of a method which can increase both contrast and resolution of an image dataset, but should not be used for quantitative intensity measurements.

An emerging field of processing methods for fluorescence microscopy images are deep learning based methods. Such methods typically require training a neural network with pairs of high-quality and low-quality images of the same field of view; the network attempts to 'learn' what series of image processing operations should be applied to reliably convert low-signal images into images closely matching the high-signal equivalent. New low-quality images (without a high-quality equivalent) can then be provided to the trained neural network, and the network will output a high-quality prediction. Example applications of these algorithms are for increasing the signal-to-noise ratio of low-signal images[27] and increasing resolution of images[28], among others[13]. Because these methods impact image intensity in a non-linear manner (**Figure 4, 'CARE'**), it is strongly recommended that intensity-dependent quantification is not performed on images processed with deep learning methods.

## Morphology

Loosely characterised as the visual appearance in terms of form or structure, morphology is critical in many biological processes because it reflects and influences the physiological state of living systems[29,30]. The vast majority of microscopy data, regardless of the modality, hold visual information that pertains to morphology. Being able to extract this type of information from image data is therefore of utmost interest, as exemplified by the popularity of software dedicated to this task such as CellProfiler[31]. Although intuitively understood by everyone, morphology is challenging to define precisely. The first challenge towards identifying the most appropriate strategy to quantify it is therefore to determine what kind of visual information is relevant. The shape of objects of interest, as for example labelled with a membrane marker, is a common readout in fluorescence microscopy[32]. In other modalities such as brightfield and electron microscopy, both shape and ultrastructure (texture) information is available[33,34]. Although the type of image feature informing on morphology may vary (whether edges, textures, or a mix of both), most morphology measurements are extracted for individual biologically-relevant objects. They therefore share the need for segmentation upstream of the actual quantification step.

### Segmenting and capturing morphology

Segmentation is the process of partitioning an image into different regions, whether background and foreground (semantic, see Glossary in **Table 1**) or individual objects (instance, see Glossary in **Table 1**). Segmentation, whether instance or semantic, is a challenging problem to solve because of the diversity of visual appearances in microscopy data. The fundamental nature of this challenge has however led to the development of numerous solutions which are available to reuse and adapt, both relying on classical image processing methods and leveraging recent machine learning tools[11]. Segmentation algorithms generally output a so-called "mask" (see Glossary in **Table 1)**, which consists of labels for each pixel in the original image (**Figure 5**). Such masks can either be binary in the case of semantic segmentation, meaning that pixels (or voxels in the case of 3D volumes) are either labelled 0 (background) or 1 (foreground) (**Figure 5b**), or composed of integer numbers for instance segmentation, whereby all pixels (or voxels) labelled with the same integer value belong to the same object instance (**Figure 5c**). Alternatively to masks, instance segmentation algorithms can also output object outlines or surfaces. In 2D images, each individual object is then identified by the list of 2D coordinates of the pixels composing its contour (**Figure 5d**). In 3D volumes, surfaces can either be represented as a list of 3D voxel coordinates, or as a more structured set of vertices and faces called mesh (see Glossary in **Table 1**). Contour (outlines) or surfaces and mask representations of individual objects can easily be converted into one another by filling the former and finding the boundaries of the latter using classical image processing methods such as connected components or boundary tracing (e.g., the marching cubes algorithm). The nature of morphology to be quantified for each biological question (e.g., colony vs individual, purely shape vs mix of shape and texture) determines the kind of readout one needs, which will in turn inform on the choice of algorithm. If individual objects are not needed, then a binary mask is sufficient. If there is no texture information, then individual contours or surfaces are sufficient. While segmentation is a necessary step towards the quantification of morphology, it is a means but not the end. The output of segmentation will be used as a basis to quantify morphology. This is worth keeping in mind to assess the level of accuracy needed from segmentation: the more subtle the morphology to be quantified is, the more accurate the segmentation must be (**Figure 3, mitochondria**). Conversely, large morphological properties such as object size may not require segmentation accuracy to the single pixel or voxel (**Figure 3, nucleus**). The scale of the morphological readout of interest thus also informs on how precise the segmentation must be for it to be quantitatively captured.

Extracting and interpreting morphology measurements

The morphology of an object can be quantified by a series of features that are collectively referred to as a feature vector (see Glossary in **Table 1**). There are two broad strategies used to construct these feature vectors. One is based on handcrafted approaches, where the feature collection is composed of measurements that are predefined by the experimenter, whilst the second strategy uses data-driven approaches, where the features list are generated through a machine learning process. A subset of commonly used handcrafted features in 2D is listed in **Table 2**. Some of these metrics are adapted from general concepts of geometry and others have been carefully engineered relying on image processing tools. All have been designed to quantify the geometrical (for shape) or visual (for texture) nature of an object in an intuitive and interpretable manner, and several can be directly extended to 3D. Different features capture different aspects of morphology, sometimes with very subtle differences (e.g., roundness and circularity).[35–38]

Collections of handcrafted features are assembled into large feature vectors to empirically capture as many aspects of morphology as possible. However, when feature vectors are built in that way, they often end up having strong correlated elements. This is due to the fact that different handcrafted measurements may be directly related to one another (for instance roundness and compactness, **Table 2**) or may be derived from the same geometrical properties (for instance area, perimeter, and circularity). Feature selection methods such as the Fisher score can be used to limit redundancy, prune the collection and retain only its most informative elements[39].

An alternate route is to let machines learn morphology descriptors directly from the data. This is relevant in many cases, from situations where morphology is too ambiguous to make it possible to craft a relevant set of features, to cases where the biological phenomenon of interest is too poorly understood to allow predicting which features are discriminative. When used well, machine learning can produce descriptors of morphology that are less biased and better capture information than manually designed ones, at the expense of interpretability and for a more significant computational cost. Features can for instance can be learned directly from flow cytometry images, addressing disagreements among experts as to which morphological readouts are able to assess red blood cell quality[40] following a weakly-supervised strategy: a deep neural network is trained to learn "good" features to describe morphology by exploiting available metadata information that relate to experimental conditions, in that case the storage date of the samples. Since organic tissues degrade, morphology is affected by sample storage time. The features learned from the image data aggregate all available morphological information to be able to describe the quality of the samples. More recent methods reduce the need for external information further by relying on concepts of self-supervised learning, in which deep neural networks only exploit information available in the input images provided for training. Learned descriptors of the morphology of single cells in fluorescence microscopy images are observed to be richer than handcrafted ones as they exploit visual information present in different fluorescence channels[41]. Given the appearance of structural markers (nucleus, cytoskeleton), a deep neural network is trained to predict the visual appearance of a protein of interest and, through this, learns a representation of overall cell morphology. The idea of learning unbiased morphology descriptors can be extended further. For example, Zinchenko et al.[42] use a deep neural network to learn descriptors of individual cell type morphology based on 3D shape and texture alone from EM volumes, without specifying a biological question. The training is here again self-supervised: an specific type of deep neural network (autoencoder[43]) learns to produce a representation that is rich enough to be able to reconstruct the input image volume while bringing together samples that are morphologically similar and pushing apart samples that are not. Many more machine learning strategies can be adopted to learn a representation of morphology from the data. Despite the effectiveness of deep neural networks, it is usually not possible to reverse-engineer the exact nature of the morphological features they rely on, making learned representations potentially difficult to interpret. Efforts to investigate and compare published approaches on benchmark or reference datasets are invaluable to navigate these available options[44].

## Factors impacting morphological quantification

It should be noted that when making quantitative measurements of sizes of objects within images, it is critical to consider both the pixel size and resolution of the image, as these provide information on the lower limits of distances that can be extracted from data (**Figure 3**). This is especially important if any measurement approaches the resolution limit of the acquired image. For example, any measurements of size should not have a dimension smaller than the theoretical resolution limit. If many objects in the image are measured to have sizes comparable to the resolution limit of the system, then this may be a population of objects of varying sizes smaller than what can be resolved. One should also remember that many morphological measurements are computed on 2D projections of structures that are actually tri-dimensional, as for instance in widefield fluorescence (**Figure 1a**) and projected confocal stacks (**Figure 1c**). These images do not take into account the third dimension and may therefore be misleading when quantifying morphology.

It is important to keep in mind that the morphology we observe in a microscopy image is a product of both the sample preparation and imaging process. Any measurements extracted to describe it are therefore strongly influenced by factors that may not be immediately relevant or obvious to the biological phenomenon of interest. For instance, some proteins commonly used as organelle markers can demonstrate apparently normal organelle morphology whilst other proteins can reveal aberrant morphology[45]. Image processing also impacts morphology measurements, as nonlinear operations can significantly alter the results of automatic thresholding, for example (**Figure 4b**). Keeping in mind which source of visual information served to construct morphology measurements is therefore crucial, in order to identify situations when one can meaningfully compare morphology readouts across different datasets. Whenever absolute image intensity is involved, for instance when relying on texture descriptors that are not only based on relative variations of intensity, one must question whether intensity can reasonably be compared across different images, as discussed in the **Image Intensity** section. Similarly, the observed shape depends on the image resolution in x, y and z, and can be strongly affected by how biological samples have been prepared for imaging. In EM for instance, each TEM will be technically specified to provide resolution in the angstrom range but the ultimate resolution of the acquired images - what can actually be visually resolved - are impacted by the sample, heavy metals introduced, thickness and density of the sample and image acquisition parameters. Introduction of significant amounts of heavy metals, may coat ultrastructural features thickly and make it difficult to resolve finer ultrastructural details, thereby limiting the possibility of quantifying morphology. Besides resolution, sample preparation is a notoriously strong factor influencing morphology. Having a good understanding of how different types of preparation distort the morphology of samples therefore provides crucial information on whether measurements can be considered biologically relevant or not. Electron microscopy has a long-standing history of investigating the effect of sample preparation[46], with examples specifically focusing on morphology preservation[47]. The need for strategies that minimally alter the structure of the imaged sample has inspired several modern fixation techniques[48,49]. As always, optimization is required to find a sensible balance of all aspects of the experimental design from sample preparation to quantification, with the ultimate aim to address the research question in mind.

# Counts and labels

Intensity and morphology can be considered "first order measurements" as they focus on quantifying purely visual information. In contrast, "second order measurements" such as counts and categorical labels (see Glossary in **Table 1**) focus on aggregating and combining morphology and intensity metrics to quantify structures that are externally defined. Counts refer to readouts enumerating the occurrences of a given structure or object in the image, while labels refer to those assigning them a category.

## Extracting and interpreting labels and object counts

Individual object count is freely obtained whenever objects can be segmented. Categorical labels can be retrieved from morphology and intensity measurements extracted from individual objects and combined in a feature vector through a classification or clustering process, depending on whether annotated examples of categories are available. However, if no other readouts are required, counting and labelling can just involve a detection process and do not necessarily require the definition of precise object boundaries (and therefore segmentation). Because of its "second order" nature, counting exploits known information of the structure or object to be detected. This can take the form of strong geometrical prior, as exploited by the Hough transform[50], or a curated example of the object of interest, as is the case in template matching[51,52]. The Hough transform, a classical image processing algorithm designed to detect occurrences of perfect circles in an image, has been successfully adapted to identify nuclei and study division in live-cell fluorescence microscopy of fission yeast[53]. Template matching can be tuned to detect an object of choice (the "template"), and is the preferred method to identify molecular complexes in EM tomograms[54,55]. Both the Hough transform and template matching are examples of algorithms that have the ability to provide object counts without going through a segmentation step.

When visual appearance varies so significantly that a single good object representative is hard to identify, deep learning methods can learn to detect occurrences of complex structures from large collections of visual examples[56,57]. Although initially designed for the detection of highly structured objects from natural images such as cars and human faces, the same algorithms have shown to generalise enough to provide good enough results in fluorescence microscopy data to allow counting[58].

At the extreme, labelling may neither require segmentation nor even object detection. Classification can be successfully carried out from tiles, obtained by splitting an image into a square grid[59]. Labels are then assigned to each tile, thus providing a readout of the categories present in the image without relying on the individually-defined objects. This approach is successfully exploited in digital pathology, where object segmentation is particularly challenging[60,61].

## Aspects impacting labels and count information

The number of elements present in an image or their category are seemingly absolute measurements, and it is thus reasonable to expect these readouts to be comparable across microscopy data. It is however important to keep in mind that, due to their "second order" nature, count and label measurements ultimately rely on morphology and intensity features. As such, when comparing across images, one should carefully consider how the nature of the data may reflect on morphology and intensity measurements (see previous two sections) and, in turn, influence the results of counting or labelling quantification pipeline.

## Quality control and confidence

While the above sections discuss different types of measurements that can be made from microscopy data, and how to interpret such measurements, the limiting factor will always be the quality of the original image being analysed. A useful phrase to bear in mind, which is used frequently in computer science, is 'garbage in, garbage out'; no matter how well-designed the analysis component of a microscopy experiment, if the images being input have poor quality, or the sample preparation and labelling have been poorly designed or executed, then the results obtained from analysis will have little meaning. In fluorescence microscopy, the most commonly-used metrics for assessing image quality are signal-to-noise ratio (SNR) and spatial resolution. SNR values alone are usually insufficient to tell whether an image will be good enough for quantitative analysis or not. Spatial resolution measurements are not necessarily an indicator of image quality directly, but can be useful for contextualising morphological measurements. For example, any measurements that have a dimension smaller than the measured resolution of the image can be attributed to noise and discarded. Similarly, if there are a large number of measurements clustered at the resolution limit then this is an indicator that it may be necessary to use a higher resolution method to study the structure of interest. Measuring image properties such as SNR and resolution, and recording them alongside other metadata from image acquisition helps to add additional context to results from quantitative analysis.

It is also important to recognise and reduce bias in quantitative image analysis. One major avenue for this is investing time in creating automated analysis pipelines whereby batches of images acquired under different biological conditions can be analysed in the same manner, free from any user input. For such automated pipelines however, it is crucial that images are acquired in similar ways and have broadly similar properties such that they all remain within acceptable tolerances of the analytical methods. This may not always be possible though, as different biological perturbations may inherently affect image quality (for example, in fluorescence microscopy, via an increase in autofluorescent species). Where automated analysis is not practical, or manual parameter selection is required, blinding can help reduce user bias[62]. Batch effects, defined as non-biological experimental variations that confound measurements, are a common source of bias, with possibly dramatic consequences on end results[63,64]. The influence of batch effects is further demonstrated by Shamir et al. who show that intensity and morphology measurements computed on microscopy images composed only of background signal can allow identifying different organelles[65]. As stated throughout this article, sample preparation, acquisition parameters (such as illumination intensity, magnification, exposure time, and detector gains), and experimental parameters (such as timestamps and sample id) must be recorded for each image whenever measurements are meant to be compared. Similarly, all parameters that can be kept constant should ideally remain as identical as possible over images. Batch effects can be further mitigated at the level of image data by correcting for intensity variations[19,20], or with feature normalisation[66]. A good summary of strategies to identify and correct batch effects is provided in Caicedo et al.[5].

The laboratory standard for assessing the legitimacy of a scientific analysis is quality control and performance metrics, and image quantification gets no exception from that. Although plenty of established metrics are available to assess the success of algorithms that carry out segmentation, detection, counting, and classification among many others, identifying metrics that faithfully reflect performance across datasets and use-cases remains an open challenge[67]. Although not quantitative, visual inspection remains a robust quality control strategy. This endeavour may however be highly non-trivial when dealing with high-dimensional, dynamic datasets or with rare events, and can be greatly facilitated by dedicated software tools[68]. Ultimately, the most powerful measure of quality control is reproducibility: the experimental procedures, microscope hardware specifications, image acquisition parameters, and image quantification algorithms provided in a published study should allow other researchers to recover its quantitative conclusions[69].

## Conclusions

An important message of this paper is that all quantitative readouts extracted from microscopy image data are to some extent the product of the image formation process and of the sample preparation protocol. Here, we focused on three big families of measurements: image intensity, morphology, and object counts or categorical labels. Unsurprisingly, image intensity is most critically affected by the image formation process and by data pre-processing or enhancement. For this reason, it is difficult to accurately compare intensity measurements regardless of the imaging modality, whether electron or light microscopy. Normalisation to a reference provides a way around this, but requires significant care to be done meaningfully. Morphology, unlike intensity, is challenging to define generally as it relates to shape, texture, and complex combinations thereof. Unless carried out on the entire image at once, morphology measurements require a first step of instance segmentation. Many different measurements related to morphology can be extracted from individual objects, making it challenging to know a priori which ones will be most informative. The best approach is therefore to combine a large number of them into a feature vector in an attempt to comprehensively describe each object's morphology. It is however important to keep in mind that morphology as observed in microscopy data is a direct product of sample preparation and of the imaging process, and that it may therefore be worth investigating how the protocol affects the aspects of morphology one wants to measure. Object counts and categorical labels obviously require objects to be identified and assigned, but may not necessarily need precise outlines. These types of readout can therefore often be obtained without explicitly segmenting individual objects. In addition to processes that lead to the image data, operations on the images themselves, prior to quantification, can dramatically affect measurements. One must therefore remain mindful of whether it is appropriate or not to perform some types of image processing given the readout of interest. Thresholding for instance removes "background", but background may as well hold true image intensity information, noise, or non-specific labelling.

Other key aspects of image quantification are quality control and confidence. While it is common practice to account for known distortions and aberrations introduced by sectioning and imaging in specific imaging modalities such as EM, assessing the accuracy or "success" of these corrections and their impact on downstream measurement is sometimes challenging. Identifying good metrics to assess whether a quantitative readout makes sense can be difficult, and plenty of confounders may adversely affect the extracted measurements. In addition to informing on the type of measurements that can meaningfully be extracted from image data, the essential information about image acquisition, sample preparation, and processing provided by metadata is therefore also crucial to allow randomization and mitigate batch effects at the analysis stage. When quantifying image data, taking into account the whole process leading from a biological sample to data analysis is ultimately the safest way to ensure that the correct measurements are extracted and that they are handled in a scientifically rigorous manner.

## Acknowledgements

VU is supported by EMBL internal funding. JJB was supported by MRC core funding to the MRC Laboratory for Molecular Cell Biology at University College London, award code MC_U12266B. SC is supported by a Royal Society University Research Fellowship (URF\R1\211329). ACC is funded by the BiPAS CDT at King's College London. The authors thank James Levitt of the Nikon Imaging Centre at King's College London for his support and assistance in this work. The authors thank Gautam Dey and his lab at EMBL, Heidelberg for providing *S. pombe* strains.

## Figure Legends

Figure 1

**Image formation in fluorescence and electron microscopy. a** Widefield microscopy of live S. pombe cells expressing sfGFP-tubulin. The whole sample volume is illuminated simultaneously and fluorescence is captured on a camera. **b** Confocal microscopy of the same field-of-view as **a**. Diffraction-limited laser spots are scanned (laser-scanning confocal) or swept in an array (spinning disk confocal, shown here) across the field of view; a pinhole (or array of pinholes) in the detection path prevents out-of-focus fluorescence from reaching the detector (point detector for laser-scanning confocal, camera for spinning disk). **c** Confocal slices can be acquired at a range of focal heights to form a 'z-stack'. This 3D volume can be projected to a single image by adding the images together ('Sum slices') or picking the most intense value for each pixel ('Max intensity'). **d** TIRF microscopy involves the generation of an evanescent field that only illuminates the volume of the sample within a few hundred nanometres of the coverslip. Emitted fluorescence is captured on a camera. Note, this is a different field of view than in **a**-**c**. All fluorescence scale bars = 10µm. **e** Transmission Electron Microscopy (TEM) of a thin section of an embedded sample. Electrons that are able to pass through the thin section of a sample are detected on a charge coupled device detector. Different sample preparation protocols can result in significantly different visualisations of a sample. A thin section of mitochondria within the cell is shown, prepared by either conventional aldehyde and osmium fixation, contrasting and resin embedding or by mild fixation, cryo protection, gelatin embedding and cryosectioning (Tokuyasu protocol). Images courtesy of J. J. Burden and I. J. White. **f** Scanning Electron Microscopy (SEM) with secondary electron (SE) detection of a whole sample. A focused beam of electrons is rastered across the surface of a whole sample (gold coated) and the SE generated from interactions of the primary electron beam with the top surface of the sample are detected by the SE detector, revealing surface topology. An exocytosis event on the surface of an endothelial cell is shown, where the structure of Von Willebrand factor strings released from the cell can be visualised. **g** Scanning Electron Microscopy with back scattered electron (BSE) detection of a whole sample. A focused beam of electrons is rastered across the surface of a whole sample (gold coated) and the BSE generated from interactions of the primary electron beam with the sample are detected by the BSE detector, revealing differences in atomic number. Gold labelled antibodies highlight Von Willebrand factor strings released from the cell as described in **f**. Image courtesy of Krupa Patel and Dan Cutler. The depth from which BSE can be generated is proportional to the voltage of the primary electron beam that reaches the sample, higher kV results in larger interaction volumes and impacts resolution. **h** Scanning Electron Microscopy with back scattered electron (BSE) detection of a thin section of a resin embedded sample. All EM scale bars = 500nm.

Figure 2

**Impact of illumination on fluorescence intensity measurements. a** Measured fluorescence intensities of Alexa Fluor 488 Phalloidin (measured from box 'A'), Mitotracker Red (box 'M') and DAPI (box 'D) in response to increasing LED illumination intensity in fixed BPAE cells (widefield microscopy, Plan Apo VC 60x Oil objective NA=1.4, 100ms exposure). Dashed lines indicate what a linear relationship between illumination intensity and fluorescence intensity would be. **b** Measured fluorescence intensities of NLS-GFP and Nup60-mCherry in live Schizosaccharomyces pombe cells (strain GD250 as described in Dey et al.[53]) in response to increasing laser illumination intensity (spinning disk confocal microscopy, SR HP Apo TIRF 100x AC Oil objective NA=1.49, 100ms exposure). Intensities were measured from regions of each channel above the Otsu threshold (portions of masks shown in corners of image). **c** Continuous spinning disk confocal imaging for 30 seconds of S. pombe cells expressing sfGFP-tubulin (strain AV2434, as described in Vještica et al.[70]) at either 10% or 70% 488nm laser intensity results in different photobleaching characteristics. Intensity was measured as the mean intensity above the Otsu threshold for each image. **d** Nuclear fluorescence intensities measured from live S. pombe cells expressing NLS-GFP ('Uncorrected') are a product of the true concentration of protein per nucleus and the flatness of the excitation illumination ('Illumination'). Inhomogeneous illumination can be corrected by dividing the acquired image by the illumination image (here, a homogenously-fluorescent slide). Image acquisition as in **b**. Scale bars in all panels = 10µm.

Table 1

**Glossary of key technical terms in image quantification.**

Figure 3

**Impact of acquisition parameters on quantitative measurements.** The same field of view of fixed BPAE cells stained with Mitotracker Red CMXRos (green) and DAPI (magenta) imaged with a widefield microscope with different objectives and additional optical magnifications ('Full field-of-view'). The number of cells in the FOV, theoretical resolution ($\Delta d$) (emission wavelength/2*NA), and resolution as measured using Image Decorrelation Analysis[71] are listed. Scale bars: 20x - 100µm, 40x, 60x - 50µm, 100x, 150x - 20µm. 'Nucleus' column shows a crop of the same nucleus from each magnification (larger white box in full FOV). The nucleus was segmented using Otsu thresholding after applying a 100nm Gaussian blur to the crop, with the threshold border indicated in yellow. Area, circularity ('Circ.') and roundness ('Round.') values are indicated below. Nucleus scale bars = 5µm. 'Mitochondria' columns show a crop of mitochondria staining at each magnification (smaller white box in full FOV). A line profile was drawn across the same region (between the arrowheads) and intensity profiles are plotted to in the right hand column (line averaging width of 5 pixels). Distances between prominent adjacent peaks were measured between the dashed lines, and are indicated below the images. Mitochondria scale bars = 1µm.

Figure 4

**Effect of image processing on downstream quantification. a** Images of a fixed BPAE cell with mitochondria stained. 'Raw' image has not been processed following acquisition; other images are processed versions of this image via the methods stated. Inset corresponds to white rectangle in the left image. Scale bars: 10µm (large image), 5µm (magnified inset). **b** Mitochondria segmented from the inset images using Otsu thresholding. Note that morphological analysis of these segmentations would yield different results between different image processing methods. **c** Histograms of pixel values within the large images in **a**, where count refers to the number of pixels. **d** For each pixel in the processed images, the pixel value is plotted against the value of the corresponding pixel in the raw unprocessed image. The grey line indicates a 1:1 relationship between processed and unprocessed pixel values (i.e. no change following processing).

Table 2

**Common handcrafted morphology features in 2D.** Descriptors based on an object's **a** shape (object geometry), **b** texture (image intensity), **c** either shape or texture.

Figure 5

**Outputs of segmentation algorithms. a** A spinning disk confocal image of S. pombe cells expressing the nuclear marker NLS-sfGFP (strain AV1200, Vještica et al.[70]). Scale bar = 20µm. **b** Semantic segmentation divides an image into two classes: foreground (i.e. objects of interest, black) and background (white). Such an image is also referred to as a binary mask. **c** Instance segmentation divides an image into background (black) and 'instances' of the object of interest. Each different instance here is randomly assigned a different colour. **d** Segmented objects can alternatively be represented by their boundaries rather than a solid object.

Figure 1

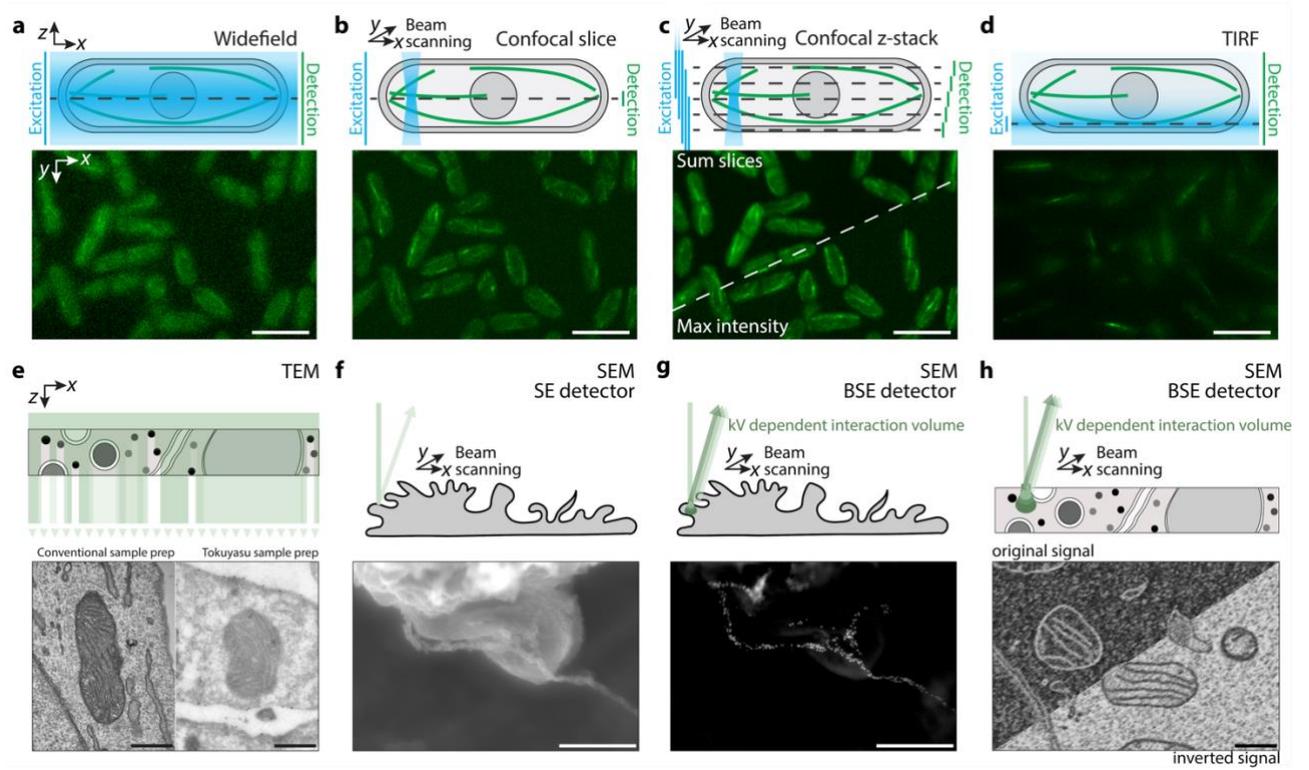

Figure 2

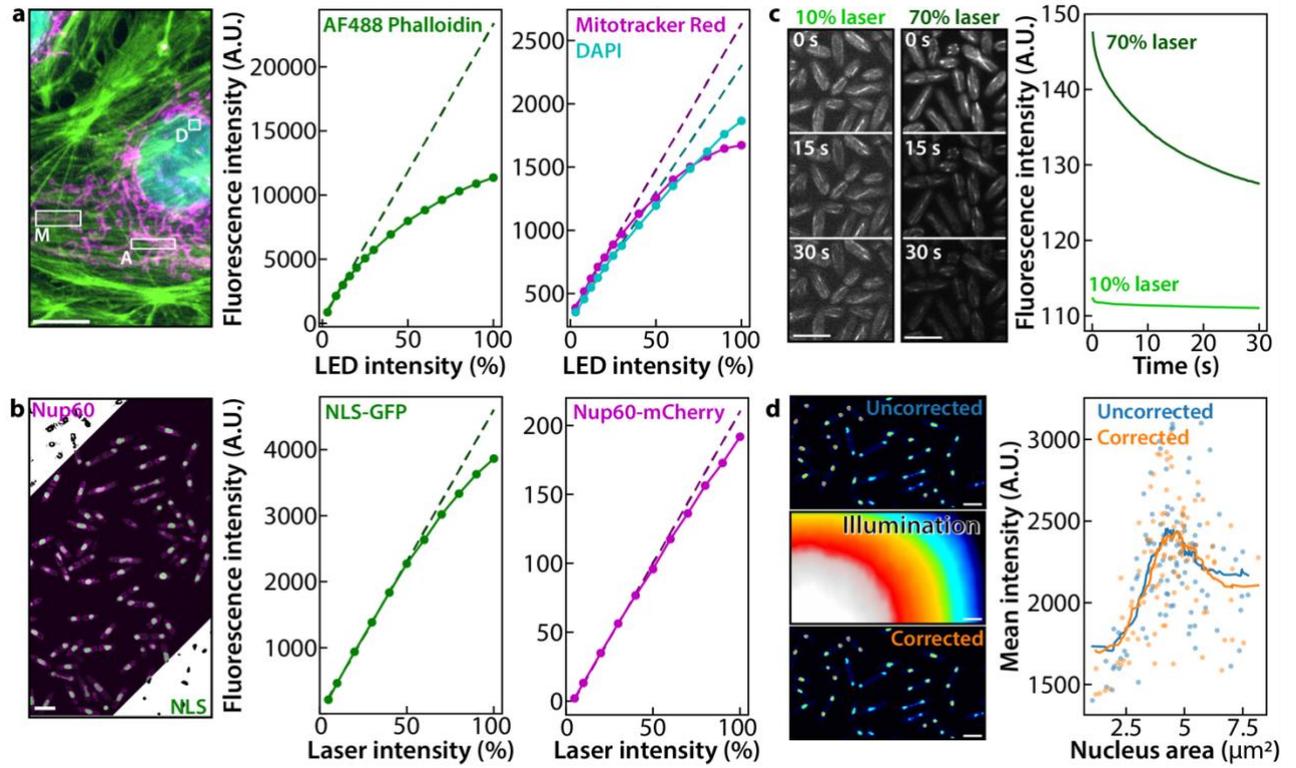

Figure 3

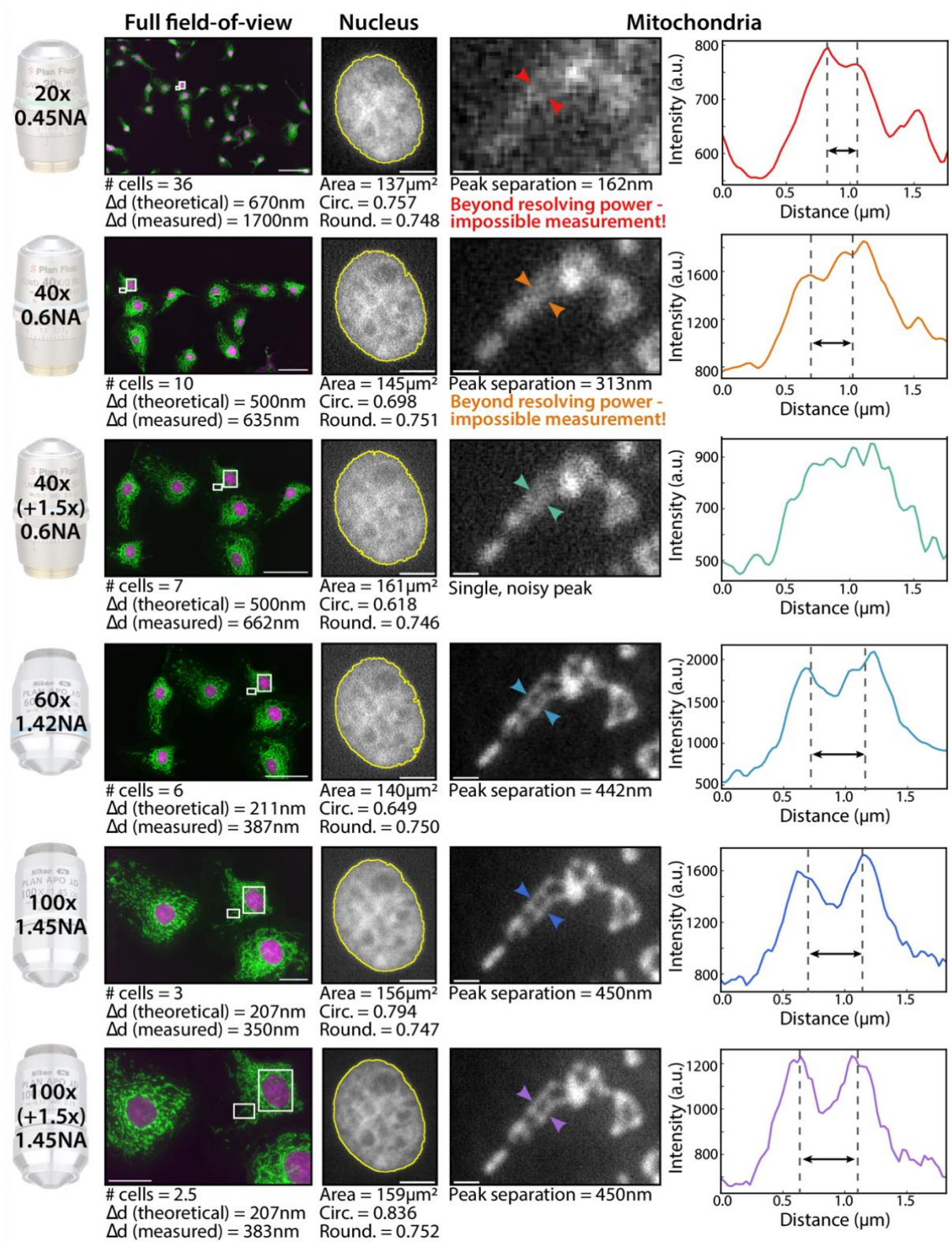

Figure 4

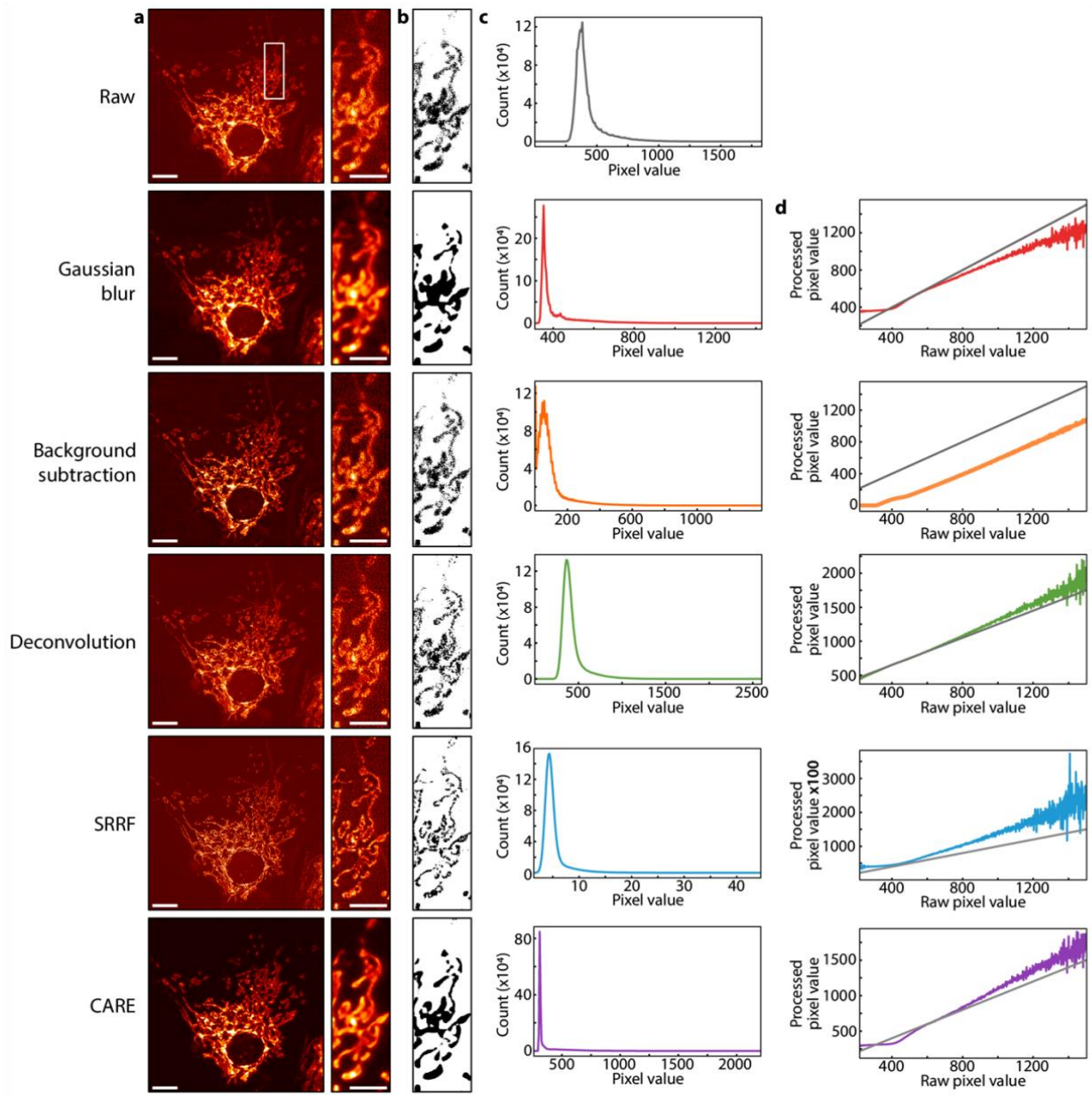

Figure 5

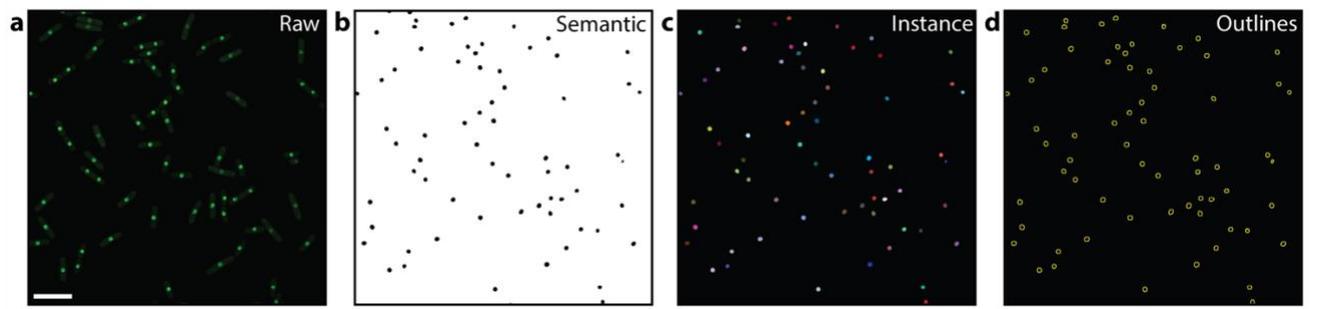

Table 1

| Glossary | |
|---|---|
| **TIRF** | (*Acquisition*) Total Internal Reflection Fluorescence. A fluorescence microscopy technique that uses total internal reflection of excitation light at the interface between the coverslip and the sample to generate a field of light that is most intense at the interface and exponentially decays with increasing depth into the sample (over a range of a few hundred nanometres). This allows for axially restricted excitation of fluorophores close to the sample. |
| **Binning** | (*Acquisition or analysis*) The process of combining the output of adjacent pixels to increase signal, thereby losing resolution. |
| **Gain** | (*Acquisition*) An amplification factor applied to the readout from the photons/electron detector in order to produce the image. It adjusts the sensitivity of the camera, but also amplifies the noise. |
| **Offset** | (*Acquisition*) The minimal intensity captured by the photons/electron detector. |
| **Pipeline** | (*Analysis*) A series of data processing steps that allows extraction of quantitative metrics from raw image data. |
| **Deconvolution** | (*Analysis*) The computational process of enhancing image contrast using knowledge of the way the microscope forms images. |
| **Semantic** | (*Analysis*) Semantic, in the context of segmentation, describes the association of each pixel of an image with a label, typically "foreground" or "background". |
| **Instance** | (*Analysis*) Individual occurrence of an object type. For example, an image with 3 circles has 3 instances of a "circle" object. In microscopy, "instances" often correspond to specific biological structures. |
| **Mask** | (*Analysis*) Image in which all pixels/voxels that are part of the foreground are set to an integer value (e.g., 1 or 255), and all pixels/voxels that are part of the background are set to 0. |
| **Mesh** | A set of vertices and faces that define polygons (often triangles) and, when taken together, form a surface covering of a 3D object. |
| **Feature vector** | A list of numbers used to quantitively represent an object. |
| **Categorical label** | (*Analysis*) An identity that is defined based on a limited and usually fixed set of possibilities (e.g., "mitochondria" or "nuclei") |

Table 2

a

| Feature Type | Metric | Definition | Interpretation | |
|---|---|---|---|---|
| Shape 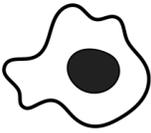 | Area | $A$ | 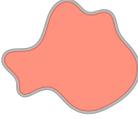 | Size of the region occupied by the object |
| | Perimeter | $L$ | 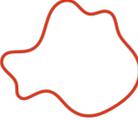 | Length of the contour of the object |
| | Feret diameters (caliper) | $F_H$ (largest) $F_V$ | 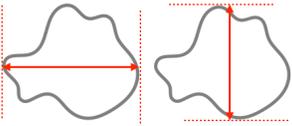 | Distances between parallel tangents touching two opposite sides of the object |
| | Convex Hull | $C$ | 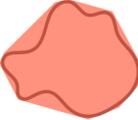 | Smallest convex shape that contains the object |
| | Circularity | $\dfrac{4\pi A}{L^2}$ | 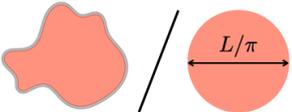 | Ratio of the object area to that of a circle with the same perimeter |
| | Roundness | $\dfrac{4A}{\pi F_H^2}$ | 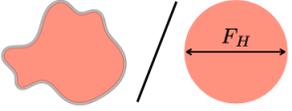 | Ratio of the object area to that of a circle with the same width |
| | Compactness | $\sqrt{\dfrac{4A}{\pi F_H^2}}$ | 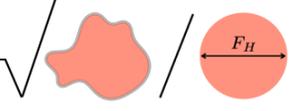 | Square-root of roundness |
| | Aspect ratio | $\dfrac{F_H}{F_V}$ | 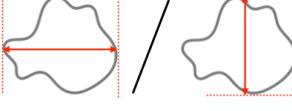 | Ratio of the object height to its width |
| | Solidity | $\dfrac{A}{A_C}$ | 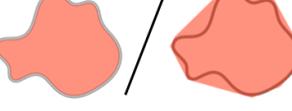 | Ratio of the object area to the area of its convex hull |

| | Convexity | $\frac{L_C}{L}$ | 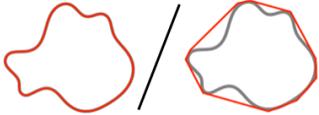 | Ratio of the convex hull perimeter to the object perimeter |

**b**

| Feature Type | Metric | Definition | Interpretation | |
|---|---|---|---|---|
| Texture | Haralick features | Defined in [34] | 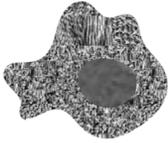 | Statistics of a matrix counting the co-occurrence of neighboring intensity values in the image (Gray Level Co-occurrence Matrix) |
| | Gabor filters | Defined in [35] | 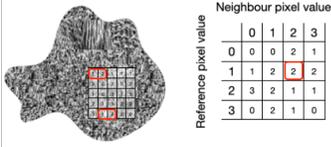 | Profile of image intensity distributions at different frequencies and orientations |

**c**

| Feature Type | Metric | Definition | Interpretation | |
|---|---|---|---|---|
| Mixed | Zernike moments | Defined in [36] | 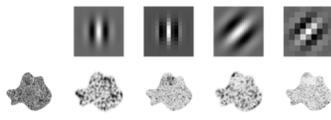 | Decomposition of the shape or texture into a basis of polynomials that are orthogonal on the unit disk |
| | Fourier descriptors | Defined in [37] | 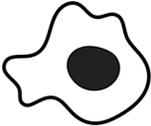 | Decomposition of the shape or texture into the Fourier basis |